# Role of Zirconium Conversion Coating in Corrosion Performance of Aluminum Alloys: An Integrated First-Principles and Multiphysics Modeling Approach

Arash Samaei [a], Santanu Chaudhuri [a, b*]

[a] Civil, Materials, and Environmental Engineering Department, University of Illinois at Chicago, IL, 60607, United States

[b] Argonne National Laboratory, Lemont, IL, 60439, United States

[*] Corresponding author's email address: santc@uic.edu

## Abstract

A variety of chromate-free conversion coatings are being actively investigated to improve the corrosion performance of light-weight alloys for aerospace and defense applications. Advancing conversion coating, however, requires an in-depth understanding of the underlying corrosion mechanisms in order to rationally design sustainable coatings. Here, we present a multiscale modeling approach to predict corrosion performance of metallic materials, with a focus on localized corrosion of Cu-containing aluminum alloys coated with $ZrO_2$ layer. First-principles and transition-state theory are used to implement the kinetics model, which includes electrolyte-metal interfacial reactions. The modeling framework systematically characterizes and couples multiple electrochemical and physical (e.g., transport) phenomena to explore interrelationships between pit morphology, surface chemistry, and local environment. This multiscale model can quantitatively link the corrosion rate of $ZrO_2$-coated aluminum alloys with the evolution of interfacial reactions during immersion, which is very difficult to establish using *in situ* experiments. We have evaluated the presented multiscale model using available experimental data. The rate of corrosion and pit stability were quantitatively assessed for various environmental parameters and applied potentials. Results show that Zr-based conversion coating strongly enhances the corrosion performance of aluminum alloys due to zirconium involvement in interfacial kinetics.

## Nomenclature

| | | | |
|---|---|---|---|
| $\Omega_{Al}$ | Aluminum matrix (anode) domain | $\varphi$ | Electrostatic potential |
| $\Omega_{\theta}$ | θ-phase (cathode) domain | $\vec{v}$ | Flow velocity |
| $\Omega_{P}$ | Passive oxide layer domain | $N^r$ | Total number of reactions |
| $\Omega_{e}$ | Electrolyte domain | $\omega_l^m$ | Stoichiometric coefficient of species $l$ in the $m$th reaction |
| $\partial\Omega_{e-m}$ | Solid-liquid interface | $k_f^m$ | Forward reaction rate constant |
| $E_{Elec(0K)}$ | Electronic energy at 0 K | $k_b^m$ | Backward reaction rate constant |
| $E_{298}$ | Total energy | $i_a$ | Total anodic current density |
| $G_{298}$ | Gibbs free energy | $i_{a,1}$ | Active dissolution current density |
| $H_{298}$ | Enthalpy | $i_{a,2}$ | Oxide formation current density |
| $\Delta G_{298,act}$ | Activation Gibbs free energy at 298 K | $\theta_0$ | Oxide coverage fraction |
| $\Delta S_{298,act}$ | Activation  entropy | $i_{a,P}$ | Current density of oxide dissolution |
| $\Delta H_{298,act}$ | Activation enthalpy | $\psi$ | Applied potential |
| $H_{Corr}$ | Enthalpy correction | $i^*$ | Concentration-independent part of exchange current density |
| $-TS_{298}$ | Entropy correction | $C^s$ | Species concentration around the electrode surface |
| $E_{298(0K)}$ | Zero-point energy | $\omega$ | Reaction order |
| $k_B$ | Boltzmann's constant | $E^0$ | Reversible potential |
| $h$ | Planck's constant | $\alpha$ | Electrochemical transfer coefficient |
| $T$ | Temperature | $M_m$ | Metal molecular weight |
| $R_g$ | Gas constant | $\rho_m$ | Density |
| $C$ | Species concentration | $z_m$ | Effective valence |
| $\overrightarrow{N_j}$ | Flux density for dilute solution | $\vec{n}$ | Outward-pointing unit normal vector on the electrolyte domain |
| $R_j$ | Reaction rate of reaction $j$ | $i_{c,m}$ | Metal-ion-reduction component of total cathodic current density |
| $D$ | Diffusion coefficient  is the | $C_j^\infty$ | Bulk concentration of species $j$ |
| $z$ | Charge number | $i_{e-m,j}$ | Current density resulting from an electrochemical reaction involving species $j$ |

## 1. Introduction

Light-weight aluminum alloys with excellent mechanical properties have been extensively used in defense and aerospace applications. Yet, these alloys are prone to localized pitting corrosion, which often takes place on the surface of passivated metals with a native oxide layer, due to the presence of heterogeneous microstructures [1, 2].  When aggressive anions (e.g., $Cl^-$) attack aluminum alloys, severe localized pits with sizes ranging from nano- to millimeters can form, which are difficult to diagnose and frequently initiate cracks that eventually lead to catastrophic failures [3]. The complexity of this issue is primarily due to two factors: the microstructure of the alloy and the corrosion environment, both of which vary dynamically with time and exhibit significant heterogeneity [4].

Aluminum alloys possess complex microstructures due to the presence of intermetallic particles (IMPs) with a broad range of compositions and sizes, as well as periphery phases around composite particles and clustering [5]. The IMPs, which have a variety of compositions and phases, are believed to be critical sites for the initiation of localized corrosion due to particle dealloying, etching of the alloy matrix, and Cu redistribution on the surface [6, 7]. The latter phenomena drastically alter the local electrochemical activity on the surface of aluminum alloys, hastening the progression of localized pitting corrosion [8, 9]. Extensive efforts have been made to characterize the electrochemical and dissolution behaviors, such as pitting potential, corrosion potential, and polarization curves, of all forms of IMPs (e.g., $Al_2Cu$ and $Al_2CuMg$) within aluminum alloys under various corrosion environments [10-16]. These studies demonstrated that the local corrosion environment governs the main phenomena at the alloy-solution interface, such as IMP dealloying, aluminum oxide formation and dissolution, matrix etching, and Cu redistribution. Moreover, pit morphology, electrical potential, electrolyte chemistry, and microstructure of aluminum alloys, all of which are space-time variant features, exhibit coupled effects on localized pitting corrosion. However, differentiating the effects of each individual feature on pitting corrosion through experiments is very challenging. In particular, the distinct role of galvanic coupling and electrolyte chemistry in the initiation and propagation of localized pits around copper-rich IMPs still need investigations [17].

In order to improve the corrosion performance of aluminum alloys, numerous surface modification techniques, such as organic (or polymeric), inorganic (e.g., anodization, conversion coating, physical or chemical vapor deposition, and electroplating), and hybrid organic-inorganic coatings (e.g., sol-gel coating), have been broadly investigated [18]. Conversion coating is used as a stronger pretreatment coating in a variety of coating applications. Chromate conversion coatings (CCCs) are the most important coatings used in a wide range of applications in the aircraft and defense industries due to their exceptional corrosion performance and great adhesion properties [19]. CCCs are chromium compounds in the 6+ oxidation state that provide self-healing corrosion inhibition due to the reduction of Cr(VI) within the coating to insoluble Cr(III) compounds [20]. The ability of such coatings to reform a Cr-containing protective barrier after the CCC layer is broken during the chemical process is referred to as self-healing.

Yet, recent studies have shown that hexavalent chromium poses significant environmental and health hazards [21, 22]. CCCs threat workers' life due to the high degree of toxicity and carcinogenicity of chromate compounds, thus their usage is largely prohibited within the United States (by the Occupational Safety and Health Administration) and the European Union (by the Restriction of Hazardous Substances Directive). These issues have motivated researchers to develop chromate-free conversion coatings with equal or better corrosion performance over the last two decades [23-26]. Several chromate-free conversion coatings, including rare-earth metal (REM), Ti, and Zr-based coatings, have been introduced to improve the corrosion performance of aluminum alloys. Among these alternatives, Zr-based conversion coating (ZrCC) has the advantages of having a low environmental impact while also being abundant and economically viable [27-29]. Extensive experimental efforts have been devoted to improving corrosion performance of ZrCC for aluminum alloys [24, 26, 30-34]. However, development of an effective ZrCC for long-term corrosion prevention requires a thorough understanding of the underlying interfacial kinetics and surface chemistry, as well as optimization of coating formulations and compositions. Since characterizing the above-mentioned space-time variant features is experimentally very challenging, multiscale modeling and simulations offer chemo-physics-based tools for investigating the controlling factors for corrosion prevention.

To date, various modeling approaches, including finite element method (FEM) and first-principles, have been reported to investigate corrosion of metallic materials such as steel (or iron), magnesium, and aluminum alloys [3, 4, 17, 35-51]. Aiming at energetic calculation and phase stability evaluation, ab initio was used to investigate surface phenomena such as dissolution, adsorption, and oxygen reduction reaction with metal/alloy at the nanoscale [52, 53]. Yet, such simulations are limited to very small systems that cannot properly describe the localized pitting process and dynamics related to those phenomena, all of which may take months or years. FE models, on the other hand, have been widely used for corrosion studies of different materials due to the ability to simulate localized pits with real-world dimensions and environmental conditions. However, according to Xiao and Chaudhuri [4], most of these models were developed based on simple and fixed pit geometry and simplified solution chemistry, with critical assumptions such as zero concentration gradient in electrolyte and time-invariant interfacial kinetics. Furthermore, they used simple metal compositions, which are inapplicable to aluminum alloys with heterogeneous microstructures and compositions [4, 5]. Current FE models rely heavily on experimental-based corrosion mechanism parametrization, and thus the availability of sufficient and accurate experimental data severely limits the predictability of such models.

Incorporating first-principles into a multiscale modeling framework is believed to significantly improve predictive corrosion models [4, 54]. Yet, to the best of our knowledge, no multiscale modeling has been conducted on the localized pitting corrosion of light-weight alloys coated with conversion coatings. This work develops a multiscale model that fully describes the coupled effects of electrolyte chemistry, interfacial kinetics, alloy microstructure, and pit morphology on the corrosion rate and stability of pits. Using density functional theory (DFT) and FEM, this model explores the interrelationships between these features to promote a fundamental understanding of localized pitting corrosion in aluminum alloys with Zr-based conversion coatings. The proposed kinetic model for electrolyte and surface chemistry was completed using DFT and transition-state theory. This research focuses on the propagation and repassivation of a localized pit in vicinity of a cathodic IMP. This work contributes to reducing the number of unknown parameters for underlying reaction mechanisms and hastening the development of more comprehensive models for corrosion resistance of conversion coatings. Section 2 will present a generic multiscale modeling framework with details of the DFT calculations and multiphysics model, Section 3 will discuss the results, and Section 4 will summarize the implications of this research.

## 2. Multiscale Modeling of Localized Corrosion

Figure 1 shows the system investigated in this work, which involves the exposure of an aluminum alloy to a sodium chloride (NaCl) solution. The alloy microstructure was obtained from SEM images [55] to serve as model geometry for simulations that mimics a real system. The SEM image, shown in Figure S3.3, is associated with a scribed coated aluminum alloy panel exposed to 0.01 M NaCl neutral solution for 100 hours of salt fog testing. The model considers a system comprised of a pure aluminum matrix and an IMP region, i.e., θ-phase ($Al_2Cu$). Note that this is a simulation of a damaged spot such as a scribed panel that can expose the IMPs. Two layers of metal oxides protect the Al matrix: a thin inner layer of native aluminum oxide ($Al_2O_3$) and a thick deposited layer of zirconium oxide ($ZrO_2$). This study focuses on pit propagation and repassivation near a cathodic IMP, with repassivation triggered by the formation of a passive layer of aluminum oxide and other insoluble precipitates on the surface [13, 14]. The model considers four main domains, as shown in Figure 1(a): aluminum matrix $\Omega_{Al}$ as anode, θ-phase ($Al_2Cu$) $\Omega_\theta$ as cathode, passive oxide layer (i.e., including both $Al_2O_3$ and $ZrO_2$ layers) $\Omega_P$, and electrolyte (NaCl solution) $\Omega_e$. The solid-liquid interface is defined as $\partial\Omega_{e-m} = \partial\Omega_e \cap (\partial\Omega_{Al} \cup \partial\Omega_\theta \cup \partial\Omega_P)$.

Figure 1(b) presents a generic multiscale modeling framework for investigating localized corrosion in light-weight alloys. It includes three layers of modeling blocks that are distinguished by their dependencies and relationships with one another and are linked to an experimental data block. The modeling process starts with inputs from experimental data such as pit surface geometry and corrosion environment. In the first modeling layer, first-principles and transition state theory are used to calculate required thermodynamic parameters, including internal energy, enthalpy, and Gibbs free energy of activation for electrochemical reactions. The Eyring-Polanyi relation, which will be explained in the following section, aids in obtaining rate constants for dissolution reactions in order to complete the kinetics model. The reaction kinetics serves as an input in the second layer, where the mass transport model is coupled with electrochemical reactions to characterize the chemical (i.e., material concentrations) and electrical (i.e., electrostatic potential) environments. In the third layer, the corrosive environment is used to determine both reactive species fluxes through the solid-liquid interface and changes in interface location. Subsequently, the pit boundary location and condition for the specified electrolyte model are obtained.

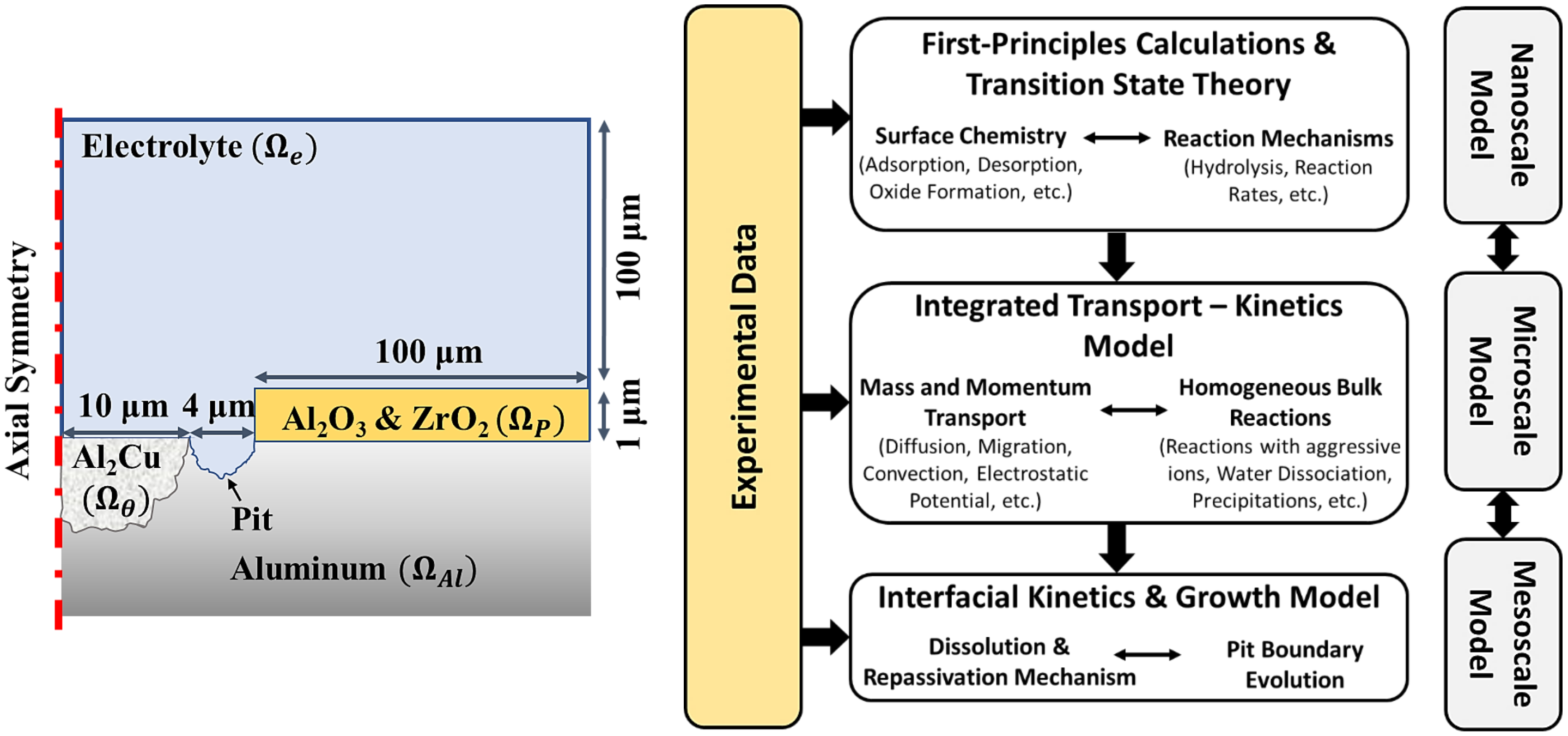

(a) (b)

*Figure 1: (a) Schematic showing the geometry used for corrosion models. A pit is located adjacent to the cathodic region, which is composed of $Al_2Cu$ IMPs ($\theta$-phase). A layer of Zr-based conversion coating covers the aluminum matrix, beneath which is a very thin layer of native alumina. The coating layer has a thickness of 1 μm. The models incorporate real-world pit geometry. The electrolyte is a sodium chloride (NaCl) solution. (b) The scheme for multiscale modeling of localized corrosion of alloys/metals. First-principles in conjunction with transition-state theory are used for energetic calculations to eventually determine the values of reaction parameters (e.g., barrier energy and rate constants). Coupling the proposed kinetics with a mass transport model, the corrosive environment over the alloy is determined. Finally, the growth of pit and its stability are quantitatively predicted.*

## 2.1 First-Principles Calculations

Among the various mechanisms for metal hydrolysis reaction in aqueous solution reported in the literature, this work used the reaction pathway for first-order hydrolysis introduced by Dong et al. [56]. The steps for calculating the energy barrier of reactions are explained as follows in this work. In the first step, the initial and final images, as their coordinates are provided in the supplementary document, were constructed. Next, density functional theory (DFT) calculations were performed to obtain the energies and forces, which are originated due to the potential energy surface (PES), acting on the system of atoms (or the images).

We used Quantum Espresso (QE) package [57] with non-conserving pseudopotential. The exchange correlation functional was approximated by generalized gradient approximation (GGA) in the formulation of PBE (Perdew-Burke-Ernzerhof) [58]. The wave function in QE was expanded through plane waves up to energy of 50 Rydberg. In the next step, climbing image – nudged elastic band (CI-NEB) method was used to determine the reaction pathways [59, 60]. This method works through converging a trial initial path in the PES to the minimum energy pathway (MEP) near the initial path [61]. The number of images used to construct the path determines its resolution. CI-NEB, despite the basic NEB method, forces one of the images to have the highest PES [61]. Finally, interpolating over converged images provided the energy barrier.

All DFT calculations were carried out at 0 K. Supplementary information includes the Cartesian coordinates of all reaction species for reactants and transition states. The electronic energies $E_{Elec(0K)}$ for aqueous reaction species were computed using DFT calculations. Subsequently, the thermodynamic variables total energy $E_{298}$, Gibbs free energy $G_{298}$, and enthalpy $H_{298}$ for those species were obtained by adding enthalpy correction $H_{Corr}$, entropy correction $-TS_{298}$, and zero-point energy $E_{298(0K)}$ onto $E_{Elec(0K)}$ [62]. The zero-point energy can be obtained from frequency calculations. $H_{Corr}$, $E_{298}$, and $-TS_{298}$ are determined using thermochemistry analyses carried out at 1 atm and 298.15 K. The Eyring equation yields the transition-state rate constant:

$$k_{TST} = \frac{k_B T}{h} \cdot \exp\left(-\frac{\Delta G_{298,act}}{R_g T}\right) = \frac{k_B T}{h} \cdot \exp\left(-\frac{\Delta S_{298,act}}{R_g T} - \frac{\Delta H_{298,act}}{R_g T}\right) \tag{1}$$

where $k_B$, $h$, $T$, $R_g$, $\Delta G_{298,act}$, $\Delta S_{298,act}$, and $\Delta H_{298,act}$ denote the Boltzmann's constant, Planck's constant, temperature, gas constant, activation Gibbs free energy at 298 K, activation entropy, and activation enthalpy, respectively.

### 2.2 Multiphysics Model

The multiphysics model used in this work is similar to one introduced and detailed previously by Xiao and Chaudhuri [4]. The following explains its key features, including certain variations in reaction kinetics compared to the one presented in [4].

#### 2.2.1 Material Balance

Both mass transport and reaction kinetics have a direct impact on the dynamical changes in mass for each chemical species $j$ in the electrolyte solution, which can be expressed mathematically as follows:

$$\frac{\partial C_j}{\partial t} = -\nabla \cdot \overrightarrow{N_j} + R_j \tag{2}$$

where $C$, $\vec{N}$, and $R$ represent the concentration ($mol/m^3$), flux density ($mol/(m^2 s)$), and reaction rate ($mol/(m^3 s)$), respectively. Since species mass transfer is driven by differences in electrical or chemical potential between two sites in an aqueous solution, the flux density $\vec{N}$ for the dilute solution can be expressed in terms of convection, migration, and diffusion:

$$\overrightarrow{N_j} = -D_j \nabla C_j - z_j F \frac{D_j}{R_g T} C_j \nabla \varphi + C_j \vec{v} \tag{3}$$

where $D$ is the diffusion coefficient ($m^2/s$); $z$ is the charge number; $R_g$ is the gas constant ($8.314\ J/(mol\ K)$); $\varphi$ denotes the electrostatic potential ($V$); and $\vec{v}$ is the flow velocity, which is zero in the case of a stagnant electrolyte solution.

The general form of the chemical production or consumption rate can be given as [63]

$$R_j = \sum_{m=1}^{N^r} \left\{ -\omega_j^m \left( k_f^m \prod_{\forall \omega_l^m > 0} (C_l)^{\omega_l^m} - k_b^m \prod_{\forall \omega_l^m < 0} (C_l)^{-\omega_l^m} \right) \right\} \tag{4}$$

where $N^r$, $\omega_l^m$, $k_f^m$, and $k_b^m$ are the total number of reactions, the stoichiometric coefficient of species $l$ in the $m$th reaction, the forward reaction rate constant, and the backward reaction rate constant, respectively. For reaction products, the stoichiometric coefficient is negative, while for reactants, it is positive. Historically, all homogeneous bulk reactions were assumed to be in equilibrium at any given position in solution and time [3, 4, 41, 45, 55]. This assumption is avoided in our work because it could lead to significant errors in calculating chemical species concentrations at high dissolution current densities [3].

The electroneutrality condition given below allows us to solve equation (2) to obtain both the electrostatic potential and the species concentration for the dilute solution:

$$\sum_j z_j C_j = 0 \tag{5}$$

### 2.2.2 Surface Reaction Mechanism

This section focuses on interfacial phenomena so that interfacial kinetics, material-solution chemistry, and surface morphology can be mathematically correlated.

#### 2.2.2.1. Anodic Kinetics

The pit state transition resulting from rivalry between passive oxide layer formation and dissolution (of aluminum in our case) is used in this multiphysics model [4, 64]. The total current density ($A/m^2$) associated with the anode can be expressed as follows:

$$i_a = (i_{a,1} + i_{a,2})(1 - \theta_0) \tag{6}$$

where $i_{a,1}$ and $i_{a,2}$ denote the magnitudes of active dissolution and oxide formation (of aluminum in this study) current densities, respectively. $\theta_0$ is the oxide coverage fraction:

$$\theta_0 = \frac{i_{a,2}}{i_{a,2} + i_{a,P}} \tag{7}$$

in which $i_{a,P}$ is the current density of oxide dissolution. Based on Anderko et al. [65], adsorption of aggressive and inhibitive species is respectively responsible for active dissolution and oxide formation (of aluminum). The current densities in equation (6) can be quantitatively correlated to temperature $T$, electrostatic potential $\varphi$, applied potential $\psi$, electrolyte chemistry, and materials as expressed below [4, 64]:

$$i_{a,1} = \sum_j i_{1,j}^* \left(C_j^s\right)^{\omega_j} \cdot \exp\left(\frac{\alpha_j F\left(\psi - \varphi - E_{M,j}^0\right)}{R_g T}\right) \tag{8}$$

$$i_{a,2} = \sum_j i_{2,j}^* \left(C_j^s\right)^{\omega_j} \cdot \exp\left(\frac{\alpha_j F\left(\psi - \varphi - E_{MO,j}^0\right)}{R_g T}\right) \tag{9}$$

where $i^*$, $C^s$, and $\omega$ are the concentration-independent part of exchange current density [64], species concentration around the electrode surface [65], and reaction order, respectively. $E^0$ and $\alpha$ are the reversible potential and electrochemical transfer coefficient, respectively. The subscript $j$ represents an aggressive ion for $i_{a,1}$ and an inhibitive ion for $i_{a,2}$. The reaction kinetics, which is strongly dependent on the solution chemistry, can be written as:

$$i_{a,P} = \sum_j k_{P,j} \left(C_j^s\right)^{\omega_j} \tag{10}$$

The corrosion rate can be obtained from the following relationship:

$$R_{Corr} = \frac{M_m}{z_m F} \frac{\int_{\partial\Omega_{e-a}} i_a dS}{\int_{\partial\Omega_{e-a}} dS} \tag{11}$$

in which $\partial\Omega_{e-a} = \partial\Omega_a \cap \partial\Omega_{e-m}$.

#### 2.2.2.2 Cathodic Kinetics

The general format of the kinetic expression for electrochemical reduction reactions on cathodes is as follows:

$$i_c = -\sum_j i_{c,j}^* \left(C_j^s\right)^{\omega_j} \cdot \exp\left(-\frac{\alpha_j F\left(\psi - \varphi - E_j^0\right)}{R_g T}\right) \tag{12}$$

where the subscript $j$ represents a dissolved ion to be reduced.

#### 2.2.2.3 Interface Boundary Evolution

The following relation can be used to track the position of the solid-liquid interface, $\Lambda$, during pit growth:

$$\frac{\partial\vec{\Lambda}}{\partial t} = -i_{e-m,m} \frac{M_m}{\rho_m z_m F} \vec{n} \qquad \forall\vec{\Lambda} \in \partial\Omega_{e-m}(t) \tag{13}$$

where $M_m$, $\rho_m$, $z_m$, and $\vec{n}$ denote the metal molecular weight, density, effective valence, and outward-pointing unit normal vector on the electrolyte domain, respectively. $i_{e-m,m}$ can be determined using the following function, which is time and location dependent:

$$i_{e-m,m} = \begin{cases} i_a & \forall\vec{\Lambda} \in \partial\Omega_a(t) \cap \partial\Omega_{e-m}(t) \\ i_{c,m} & \forall\vec{\Lambda} \in \partial\Omega_c(t) \cap \partial\Omega_{e-m}(t) \\ 0 & \forall\vec{\Lambda} \in \partial\Omega_P(t) \cap \partial\Omega_{e-m}(t) \end{cases} \tag{14}$$

where $i_{c,m}$ denotes the metal-ion-reduction component of total cathodic current density [4]. According to Equations (13) and (14), $i_a$ determines the speed at which the pit propagates through the anode. There is no material transfer between the oxide layer ($ZrO_2$) and electrolyte solution. Note that we have a native oxide passivation mechanism for alumina to recalculate the exposed metal surface and oxide covered area in the pit.

#### 2.2.2.4 Initial and Boundary Conditions

The multiphysics model is based on mass transport phenomena in electrolytes with moving pit boundaries. In this study, the following relationships were considered for the initial conditions:

$$C_j(\vec{\Lambda}, t = 0) = C_j^{\infty} \qquad \forall\vec{\Lambda} \in \Omega_e \tag{15}$$

$$\varphi(\vec{\Lambda}, t = 0) = \varphi^{\infty} \qquad \forall\vec{\Lambda} \in \Omega_e \tag{16}$$

in which $C_j^{\infty}$ and $\varphi^{\infty}$ denote the bulk concentration of species $j$ and the electrostatic potential, respectively. Other conditions that can be applied to the boundaries far from the pit are as follows:

$$C_j(\vec{\Lambda}, t > 0) = C_j^{\infty} \qquad \forall\vec{\Lambda} \in .\partial\Omega_e/\partial\Omega_{e-m} \tag{17}$$

$$\varphi(\vec{\Lambda}, t > 0) = \varphi^{\infty} \qquad \forall\vec{\Lambda} \in .\partial\Omega_e/\partial\Omega_{e-m} \tag{18}$$

While the initial boundary $\partial\Omega_{e-m}(t = 0)$ is associated with the surface morphology obtained from experiments, equations (13) and (14) determine the boundary location $\partial\Omega_{e-m}(t > 0)$ during the corrosion process. The flux of chemical species involved in interfacial reactions can be obtained using the following relationship:

$$\vec{N_j}(\vec{\Lambda}, t > 0) = \frac{-i_{e-m,j}}{z_j F}\vec{n} \qquad \forall\vec{\Lambda} \in \partial\Omega_{e-m} \tag{19}$$

where $i_{e-m,j}$ denotes the current density resulting from an electrochemical reaction involving species $j$ (see equation (14) to obtain this term).

## 3. Case Study: Aluminum Alloy with Zr-Based Conversion Coating

In this work, the presented multiscale modeling approach was used to investigate localized pitting corrosion in aluminum alloys coated with a zirconia layer and immersed in NaCl solution (Figure 1(a)). To better understand the critical phenomena that occur during corrosion, some assumptions were considered in order to simplify the current problem. The following are the assumptions: 1) dealloying of $Al_2Cu$ is inhibited by covering a passive layer on the θ-phase, as found in experiments [12], 2) Cu-plating on the θ-phase is ignored after covering this phase with an insoluble CuCl passive layer, 3) the zirconia layer (i.e., $\Omega_P$) remains intact during the simulations, and 4) $Al(OH)_3$ precipitation as a corrosion product on the pit is ignored due to its low concentration when $pH < 4$ [3]. Because of hydrolysis reactions, the solution in an active pit is very acidic [66].

### 3.1 Reaction Mechanisms

The kinetics model associated with the localized corrosion of an aluminum alloy coated with a zirconia layer consists of homogeneous bulk reactions involving both Al and Zr species. Table 1 contains a list of these reactions (reactions 1-16), which were compiled from various works of literature. It should be noted that the model does not account for reactive dissolution of $ZrO_2$ in the vicinity of a pit. The coating is considered to be responsible for the excess $Zr^{4+}$. The processes in $ZrO_2$ are very complex, including multistep dissolution and hydroxide formation. It is only used for native zirconium alloys.

In anodic region, a layer of oxide film can form as a result of the following reactions, which causes repassivation of the active pit in aluminum alloys:

$$\equiv 2\text{Al} + 3H_2O \rightarrow \equiv \text{Al}_2O_3 + 6H^+ + 6e^- \quad \text{for acidic to neutral solution} \tag{20}$$

$$\equiv 2\text{Al} + 6\text{OH}^- \rightarrow \equiv \text{Al}_2O_3 + 3H_2O + 6e^- \quad \text{for alkaline solution} \tag{21}$$

On the other hand, dissolution reactions of the oxide film at different pH levels are as follows:

$$\equiv \text{Al}_2O_3 + 6H^+ \rightarrow \ 2\text{Al}^{3+} + 3H_2O \quad \text{for acidic to neutral solution} \tag{22}$$

$$\equiv \text{Al}_2O_3 + 2\text{OH}^- \rightarrow \ 2\text{AlO}_2{}^- + H_2O \quad \text{for alkaline solution} \tag{23}$$

In the cathodic region, however, oxygen reduction is assumed to be the only cathodic reaction on the Cu surface:

$$O_2 + 2H_2O + 4e^- \rightarrow \ 4OH^- \tag{24}$$

### 3.2 Simulation Inputs

The material properties used in the modeling for aluminum and zirconium are: $M_{w,Al6061} = 26.98\ g/mol$, $\rho_{Al6061} = 2700\ kg/m^3$, $M_{w,Zr} = 91.22\ g/mol$, and $\rho_{Zr} = 6506\ kg/m^3$. The following dissolved species were considered in the simulations: $Zr^{4+}$, $ZrO^{2+}$, $ZrOH^{3+}$, $Zr(OH)_2{}^{2+}$, $ZrCl^{3+}$, $Zr(OH)Cl^{2+}$, $ZrOCl^+$, $Zr(OH)_3Cl$, $ZrO(OH)Cl$, $Al^{3+}$, $Al(OH)^{2+}$, $Al(OH)_2{}^+$, $Al_2(OH)_2{}^{4+}$, $AlCl^{2+}$, $Al(OH)Cl^+$, $Al(OH)_2Cl$, $Cl^-$, $H^+$, $OH^-$, $Na^+$, and $O_2$. The diffusion coefficients for these species were collected from the literature [3, 67]. The diffusion coefficient for Zr-containing species was considered to be the same and equal to $2.36 \times 10^{-11}\ m^2/s$ [67]. The diffusion coefficient was calculated by extrapolating the curve for the variation of applied current versus time for different temperatures provided in [67].

The component was immersed in a NaCl solution at room temperature, $298.15\ K$. The concentration of the solution was considered to have a range from $10^{-4}\ M$ to $1\ M$. A bulk concentration of $60\ \mu M$ was set for the dissolved oxygen in the electrolyte [68]. The applied potentials were changed from $-1\ V$ to $-0.55\ V$ (vs. saturated calomel electrode (SCE)). Note that this study aims at investigating the full anodic polarization behavior by changing the applied potentials. Furthermore, the free corrosion case, in which the applied potentials equal the open circuit potential (OCP), is not explored.

The parameter values for Butler-Volmert equation used for the electrode kinetics were collected from literature [3, 4]. The exchange current densities were considered to be the same and equal to $10^{-5}A/m^2$ for both oxide formation and active dissolution. The current densities for passive and cathodic dissolutions were set as $10^{-2}A/m^2$ and $10^{-4}A/m^2$, respectively. The mechanism of cathodic dissolution of aluminum involves alkalinization of the oxide/solution or aluminum/solution interfaces during oxygen and water reduction, which results in chemical dissolution of the oxide or metal [69-71]. The values of applied potential and reversible potentials for aggressive species and inhibitive species were considered as $-0.5\ V$, $-0.911\ V$, and $-0.4\ V$, respectively. The reversible potential for dissolved species in cathode was set as $-0.282\ V$. In this study, the electrochemical transfer coefficient $\alpha$ is set to 0.7.

## 4. Results and Discussion

### 4.1 Evaluation of Kinetics and Multiscale Models

To complete the kinetics model for multiphysics simulations, DFT-TST calculations were carried out to determine the missing reaction constants and energy barriers for first-order reactions involving Zr. However, reaction constants for a few reactions containing Al were obtained for evaluating the calculations due to a lack of experimental data for the rate constants and energy barriers of Zr-containing reactions. Table 1 shows the results of calculations that include the forward-reaction rate constant and the energy barrier. Comparisons of computed and experimental values for the kinetic parameters indicate that for reactions containing Al, calculations overestimated the rate constants by less than 50%. However, the results of the calculations may fall within the measurement error, which has not been reported in the literature. In addition, the difference in reaction rate constants between experiments and DFT-TST calculations can be attributed to the variation of conditions under which the data are obtained [72]. The rate constants used in the simulations are a combination of data from both experiments and DFT-TST calculations, as shown in the last two columns of Table 1. Due to very expensive computations, the values for the second-order reaction rates containing Zr were assumed in this study.

To assess the multiscale models, the corrosion rates of bare aluminum and zirconium obtained from simulations were compared to the experimental results in Table 2. In general, the model is about zirconia coated Al-6061 and the Zr 705/702 are reference for the oxide layer. $Al_2Cu$ is a cathodic impurity in Al-6061, which is used in the model for driving galvanic corrosion. Note that the only metal corroding in this paper is Al-6061. The kinetic model for the bare aluminum includes reactions (4-6), (9,10), and (14-16), reported by Guseva et al. [3]. The reactions (1-3), (7,8), (11-13), and (16) are considered for the kinetic model of Zr 705/702. The parameter values input for the three materials compared in Table 2 differ only in that the reactions considered in the kinetic model differ in these cases. The simulation inputs were set to match the experimental conditions. Model results are in good agreement (almost in the same ballpark) with measurements reported in the literature. Differences in corrosion rates between simulations and experiments can be explained by differences in the conditions under which the data is obtained. Along with the results for Al 6061 and Zr 705/702, the corrosion rate of Al 6061 with Zr-based conversion coating is reported for the same conditions as bare Al 6061 to better understand the effects of Zr-based coatings on the corrosion performance of aluminum alloys. However, there is no experimental data for Al coated with $ZrO_2$ in the literature. According to the simulation results in Table 2, Zr-based conversion coating improves the corrosion performance of aluminum alloys immersed in NaCl solution, which is in line with experimental findings [19, 73-75]. The reasons will be explained in the following section.

*Table 1: Homogeneous bulk reactions proposed for the dissolution of aluminum alloy with Zr-based conversion coating immersed in chloride solution. The energy barrier ΔG and rate constants for forward reaction $K_f$ and backward reaction $K_b$ is used to compare DFT calculations performed in this work with experimental data. The rate constants selected for experiments are a combination of experimental and DFT data.*

| **Reaction Type** | **Reaction** | **DFT – Transition State Theory** | | **Experiment** | | | **Selected Rate Constants for Simulations** | |
|---|---|---|---|---|---|---|---|---|
| | | ΔG (Cal/mol) | $K_f$ | $K_f$ | $K_b$ | Ref. [a] | $K_f$ | $K_b$ |
| **Hydrolysis** | $Zr^{4+} + H_2O \leftrightarrow ZrOH^{3+} + H^+$ (1) | $1.37\times10^4$ | $4.49\times10^2$ $s^{-1}$ | - | - | [76, 77] | $4.49\times10^2$ $s^{-1}$ | $1\times10^7$ $M^{-1}\,s^{-1}$ [c] |
| | $Zr^{4+} + H_2O \leftrightarrow ZrO^{2+} + 2H^+$ (2) | $2.53\times10^4$ | $1.24\times10^{-6}$ $s^{-1}$ | - | - | [78] | $1.24\times10^{-6}$ $s^{-1}$ | $1\times10^2$ $M^{-1}\,s^{-1}$ [c] |
| | $ZrOH^{3+} + H_2O \leftrightarrow Zr(OH)_2^{2+} + H^+$ (3) | $7.81\times10^3$ | $1.01\times10^7$ $s^{-1}$ | - | - | [76] | $1.01\times10^7$ $s^{-1}$ | $1\times10^9$ $M^{-1}\,s^{-1}$ [c] |
| | $Al^{3+} + H_2O \leftrightarrow Al(OH)^{2+} + H^+$ (4) | $1.03\times10^4$ | $1.58\times10^5$ $s^{-1}$ | $1.09\times10^5$ $s^{-1}$ | $4.4\times10^9$ $M^{-1}\,s^{-1}$ | [79] | $1.09\times10^5$ $s^{-1}$ | $4.4\times10^9$ $M^{-1}\,s^{-1}$ |
| | $Al(OH)^{2+} + H_2O \leftrightarrow Al(OH)_2^+ + H^+$ [c] (5) | $6.89\times10^3$ | $4.87\times10^7$ $s^{-1}$ | - | - | [3] | $4.87\times10^7$ $s^{-1}$ | $4.4\times10^9$ $M^{-1}\,s^{-1}$ [c] |
| | $2Al^{3+} + 2H_2O \leftrightarrow Al_2(OH)_2^{4+} + 2H^+$ (6) | - | - | $10^{-2}$ $M^{-1}\,s^{-1}$ | $10^8$ $M^{-1}\,s^{-1}$ | [79] | $10^{-2}$ $M^{-1}\,s^{-1}$ | $10^8$ $M^{-1}\,s^{-1}$ |
| **Reactions with Cl− ion** | $Zr^{4+} + Cl^- \leftrightarrow ZrCl^{3+}$ (7) | - | - | - | - | [80] | 100 $M^{-1}\,s^{-1}$ [c] | 10 $s^{-1}$ [c] |
| | $ZrOH^{3+} + Cl^- \leftrightarrow Zr(OH)Cl^{2+}$ (8) | - | - | - | - | [80] | 10 $M^{-1}\,s^{-1}$ [c] | 1000 $s^{-1}$ [c] |
| | $Al^{3+} + Cl^- \leftrightarrow AlCl^{2+}$ (9) | - | - | 226 $M^{-1}\,s^{-1}$ | $75 - k_f.(c_{Al^{3+}} - c_{AlOH^{2+}})$ $s^{-1}$ | [81] | 226 $M^{-1}\,s^{-1}$ | $75 - k_f.(c_{Al^{3+}} - c_{AlOH^{2+}})$ $s^{-1}$ |
| | $Al(OH)^{2+} + Cl^- \leftrightarrow Al(OH)Cl^+$ (10) | - | - | 19 $M^{-1}\,s^{-1}$ | $5700 - k_f.c_{AlOH^{2+}}$ $s^{-1}$ | [81] | 19 $M^{-1}\,s^{-1}$ | $5700 - k_f.c_{AlOH^{2+}}$ $s^{-1}$ |
| **Product Formation** | $ZrCl^{3+} + 3H_2O \rightarrow Zr(OH)_3Cl + 3H^+$ (11) | $2.19\times10^{4}$ | $3.62\times10^{-4}$ $s^{-1}$ | - | - | [b] | $3.62\times10^{-4}$ $s^{-1}$ | - |
| | $Zr(OH)Cl^{2+} + 2H_2O \rightarrow Zr(OH)_3Cl + 2H^+$ (12) | $2.17\times10^{4}$ | $5.26\times10^{-4}$ $s^{-1}$ | - | - | [b] | $5.26\times10^{-4}$ $s^{-1}$ | - |
| | $ZrOCl^{+} + H_2O \rightarrow ZrO(OH)Cl + H^+$ (13) | $2.52\times10^{4}$ | $1.44\times10^{-6}$ $s^{-1}$ | - | - | [b] | $1.44\times10^{-6}$ $s^{-1}$ | - |
| | $AlCl^{2+} + 2H_2O \rightarrow Al(OH)_2Cl + 2H^+$ (14) | $2.43\times10^{4}$ | $5.94\times10^{-6}$ $s^{-1}$ | $4\times10^{-6}$ $s^{-1}$ | - | [82] | $4\times10^{-6}$ $s^{-1}$ | - |
| | $Al(OH)Cl^{+} + H_2O \rightarrow Al(OH)_2Cl + H^+$ (15) | $2.10\times10^{4}$ | $1.74\times10^{-3}$ $s^{-1}$ | $4\times10^{-6}$ $s^{-1}$ [c] | - | [82] | $4\times10^{-6}$ $s^{-1}$ | - |
| **Water Dissociation** | $H_2O \leftrightarrow OH^- + H^+$ (16) | $2.32\times10^{4}$ | $4.37\times10^{-5}$ $s^{-1}$ | $2.6\times10^{-5}$ $s^{-1}$ | $1.3\times10^{11}$ $M^{-1}$ $s^{-1}$ | [83] | $2.6\times10^{-5}$ $s^{-1}$ | $1.3\times10^{11}$ $M^{-1}$ $s^{-1}$ |

[a] References are associated with both reactions and experimental data for reaction constants.

[b] Reactions are proposed in this study.

[c] Assumed.

*Table 2: Corrosion rate of bare aluminum alloy 6061, bare zirconium 705/702, and aluminum alloy 6061 with Zr-based conversion coating obtained from simulations in this work and reported experimental measurements. Certain conditions for simulations and experimental measurements are also mentioned for the purpose of comparison.*

| **Metal** | **Simulations** | | **Experiment** | | |
|---|---|---|---|---|---|
| | **Condition** | **$R_{Corr}$ (mm/year)** | **Condition** | **$R_{Corr}$ (mm/year)** | **Ref.** |
| **Al 6061** | NaCl, 30 °C, pH = 6, Steady | 0.020899 | NaCl (1%), 30 - 50 °C, pH = 5.8 – 6.0, after 480 hr | 0.0305 | [84] |
| **Zr 705/702** | NaCl, 30 °C, pH = 1, Steady | 0.0005506 | NaCl (25%), 35-100 °C, pH = 1, after 504 hr | < 0.0007 | [85, 86] |
| **Al with $ZrO_2$ Coating** | NaCl, 30 °C, pH = 6, Steady | 0.002254 | NaCl, 30 °C, pH = 6.5 – 7.2 | 0.005588 ± 0.0038 | [87] |

### 4.2 Environmental Effects on the Growth of Pit

To understand the effects of chloride ions (i.e., $Cl^-$) as aggressive species on localized pitting corrosion, the change in corrosion rate of bare Al 6061, bare Zr 705, and Al 6061 coated with Zr-based conversion coating as a function of the bulk concentration of chloride ions in the solution is plotted in Figure 2(a). Here, the alloy was immersed in a neutral NaCl solution and was potentiostatically controlled with a $-0.55\ V$ applied potential. A fixed pit geometry corresponding to a particular instant of time during pit propagation was used. The figure shows that the corrosion rate of the systems increases as the bulk concentration of $Cl^-$ was increased. This is because increasing the bulk concentration of $Cl^-$ in the solution drives more chloride ions to the pit surface, resulting in increased acidity in the pit (see Figure 2(b)) and thus increased corrosion rate of the system. Experimental observations for localized corrosion of aluminum alloys indicated the same pattern [88].

Furthermore, Figure 2(a) demonstrates that Zr-based conversion coatings promote the corrosion resistance of aluminum alloys due to the involvement of Zr in interfacial kinetics. This is due to the suppression of both the oxygen reduction reaction (ORR) and anodic dissolution, with the ORR being more strongly inhibited than bare Al 6061 due to the impediment of oxygen chemisorption on IMPs [89]. It is worth mentioning that the oxygen concentrations deplete in the pit due to all chemical reactions, leading to the difficulty in repassivation (see the oxygen concentration contour plot on the Supplementary document). The corrosion rate of Zr is very low in comparison to the other two systems. The latter is due primarily to the fact that the anodic current density (as defined in the Anodic Kinetic section) is influenced by the reactions and their rates of reaction. The kinetic models of coated and bare Al differ in terms of reaction within the pit, which influences corrosion rates.

Figure 2(b) depicts the pH distribution along the pit surface, specifically close to the solid-liquid interface. With a constant pH, the pit status changes from active to passive as the concentration of $Cl^-$ in the solution decreases, which is in line with the experimental observations [90]. Note that the pH changes over the pit surface even for dilute concentration of $Cl^-$ (e.g., $10^{-4}$M and $10^{-3}$M), which cannot be observed from Figure 2(b). We specifically added Figure S2.1 in the Supplementary document to demonstrate how the pH changes over the pit surface for $Cl^{-}$ concentration of $10^{-4}$M and $10^{-3}$M. $Al^{3+}$, $AlCl^{2+}$, $Zr^{4+}$, $ZrCl^{3+}$, and $Zr(OH)Cl^{2+}$ are the main species in neutralizing the chloride ion charges in the solution. The present model assumes that 1) Al- and Zr-containing salts are sufficiently soluble during the early stages of pit growth, and 2) continuous wet environments, similar to salt fog chambers, can allow for very high concentrations of ionic species.

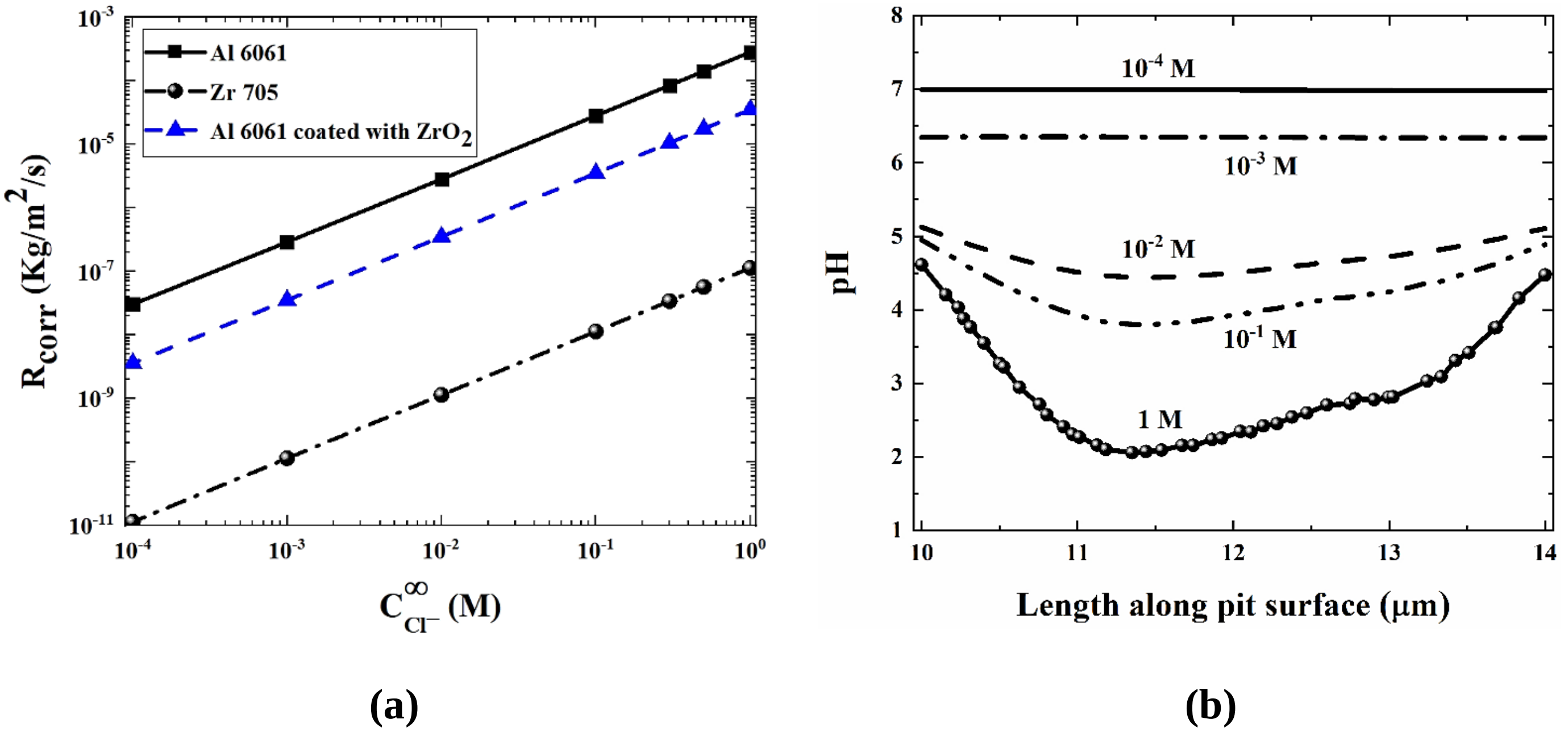

*Figure 2: (a) The effect of bulk concentration of $Cl^{-}$ on the corrosion rate of bare Al 6061, bare Zr 705, and Al 6061 coated with Zr-based conversion coating. It shows that the Zr-based conversion coating significantly improves the corrosion performance of aluminum alloys in NaCl solution. (b) Chemical conditions near the pit surface at a same preliminary time step showing how solution pH changes close to the pit surface for different concentrations of chloride ions. The curves for $Cl^{-}$ concentrations of $10^{-4}$M and $10^{-3}$M are also shown in Figure S2.1 in the Supplementary document to demonstrate how their pH changes over the pit surface.*

Figure 3(a-c) depicts the anodic polarization curves obtained by using a series of constant concentrations of $Cl^{-}$ and decreasing the applied potential $\psi$ in 10 mV steps. This plot aids in determining pit stability as the repassivation potential can be quantified using the current approach. The repassivation potential, as described by Frankel et al. [91], is the potential that the pit current density falls below a critical value, which is the minimum current density required to preserve the critical pit environment and prevent repassivation. Figure 3(a-c) shows that the corrosion rate of the systems decreases at a fixed $Cl^{-}$ concentration as the applied potential decreases from $-0.55\ V$. A sharp decrease in corrosion rate indicates pit repassivation as applied potentials are reduced. The sudden growth of an aluminum oxide layer on the metal surface accurately described pit repassivation.

Furthermore, Figure 3(a-c) demonstrates that the repassivation potential of a system (e.g., Al 6061) decreases as the concentration of chloride ions in the solution increases. It implies that the pit surface becomes less stable when surrounded by less concentrated aggressive ions (e.g., $Cl^{-}$). Our model successfully predicted the experimentally observed behavior for aluminum alloys [88, 90]. Figure 3(d) compares the polarization curves of different systems (Al 6061, Zr 705, and Al 6061 with Zr-based conversion coating) at a fixed $Cl^{-}$ concentration of $1M$. In comparison to the other two systems depicted in Figure 3(d), Al 6061 has a very high repassivation potential and a much higher corrosion rate, which is consistent with experimental observations.

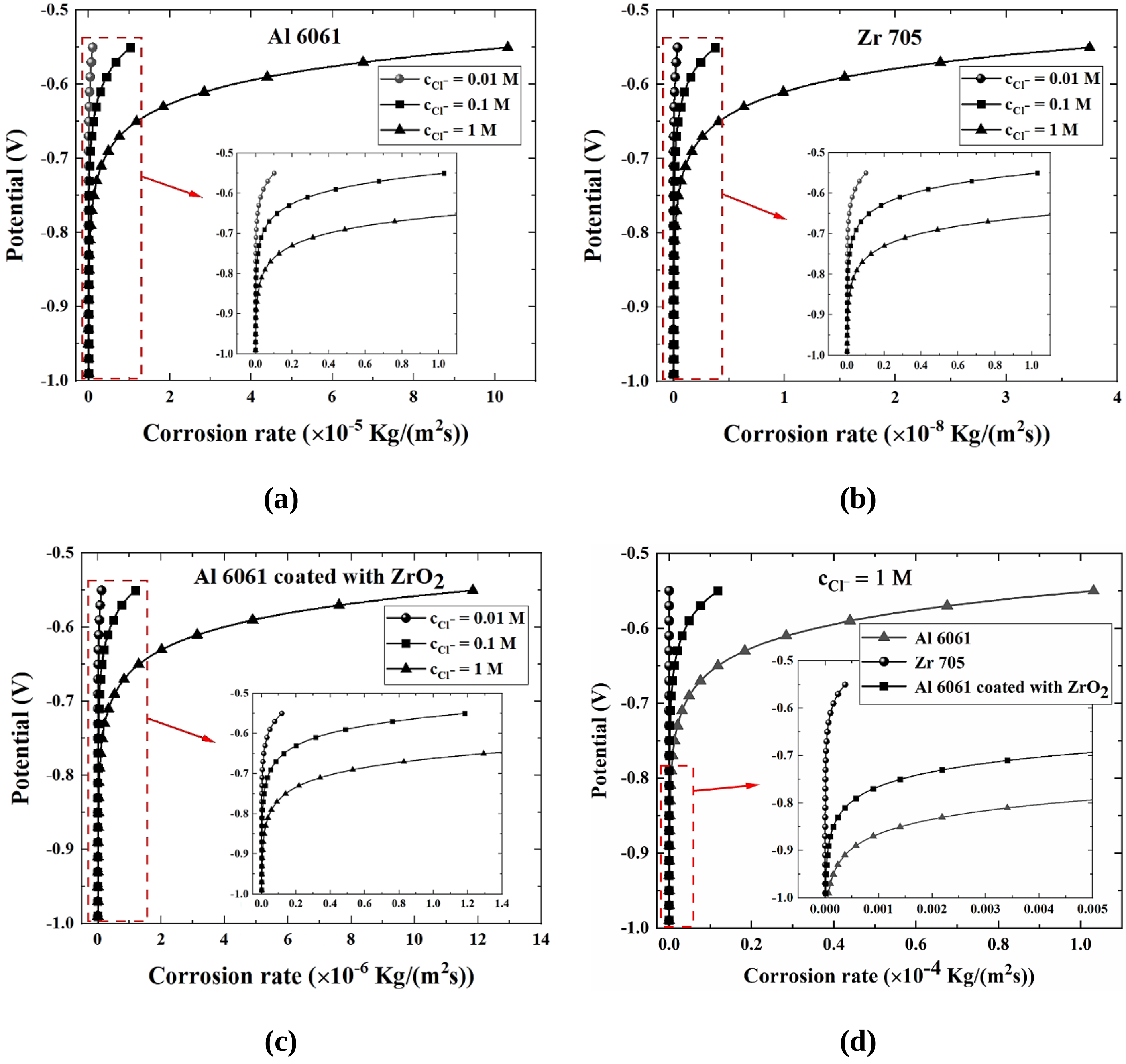

*Figure 3: Anodic polarization curves for (a) bare Al 6061, (b) bare Zr 705, and (c) Al 6061 with Zr-based conversion coating at different concentrations of $Cl^-$. It shows that the repassivation potential decreases as the concentration of chloride ions in the solution increases. The semi-log plots for these polarization curves are shown in Figure S3.2. (d) A comparison of anodic polarization curves at $c_{Cl^-} = 1M$ for different systems.*

To study the effects of pH on the corrosion rate of the systems, the concentration of chloride ions in the solution was kept constant at 0.01M. Either hydrochloric acid (HCl) or sodium hydroxide (NaOH) was used to adjust the bulk pH of the solution. Experiments have shown that when aluminum alloys are exposed to alkaline solutions, general corrosion occurs rather than localized pitting corrosion because uniform thinning of the passive oxide film (due to OH ion attack) overwhelms pitting corrosion [88]. Because localized corrosion is the focus of this research, acidic or near-neutral solutions (i.e., pH range 1 to 8) have been examined.

Figure 4(a) compares the corrosion rate of Al 6061 and Al 6061 with Zr-based conversion coating as a function of solution pH at a fixed applied potential of $-0.55\ V$. The decrease in pH results in an increase in corrosion rate, which is in line with the experimental findings [92]. According to Xiao and Chaudhuri [4], there is a critical pH above which the corrosion rate rapidly decreases to zero. Such a significant decrease shows that the pit status has changed from active to passive. The critical pH values for bare Al 6061 and Al 6061 coated with $ZrO_2$ are approximately 6 and 5, respectively, indicating that the presence of Zr changes the pit status in Al 6061 from active to passive at lower pH. The presence of Zr near the pit surface, on the other hand, indicates that pit stabilization is accelerated.

Figure 4(b) and (c) depict how the corrosion rate of Al 6061 with Zr-based conversion coating changes with pH at various applied potentials. The corrosion rate of the system increases as the solution pH decreases for different potentials due to the more acidic environment around the pit surface. At a certain pH, the corrosion rate increases as the applied potential increases. Furthermore, changes in pH between 4 and 8 have almost no effect on the corrosion rate. Figure 4(d) shows a 3D pH-potential diagram that can be used to quickly determine the corrosion rate and pit stability of Al 6061 with Zr-based conversion coating under any corrosive conditions. This 3D diagram expands on the concept of Pourbaix diagram, which shows the equilibrium phases of specific metal-electrolyte systems at various pH and potentials, by adding a new dimension of information, namely, corrosion rate. According to Figure 4(d), a more acidic environment near the pit surface results in the continuous growth of an active pit at a lower applied potential.

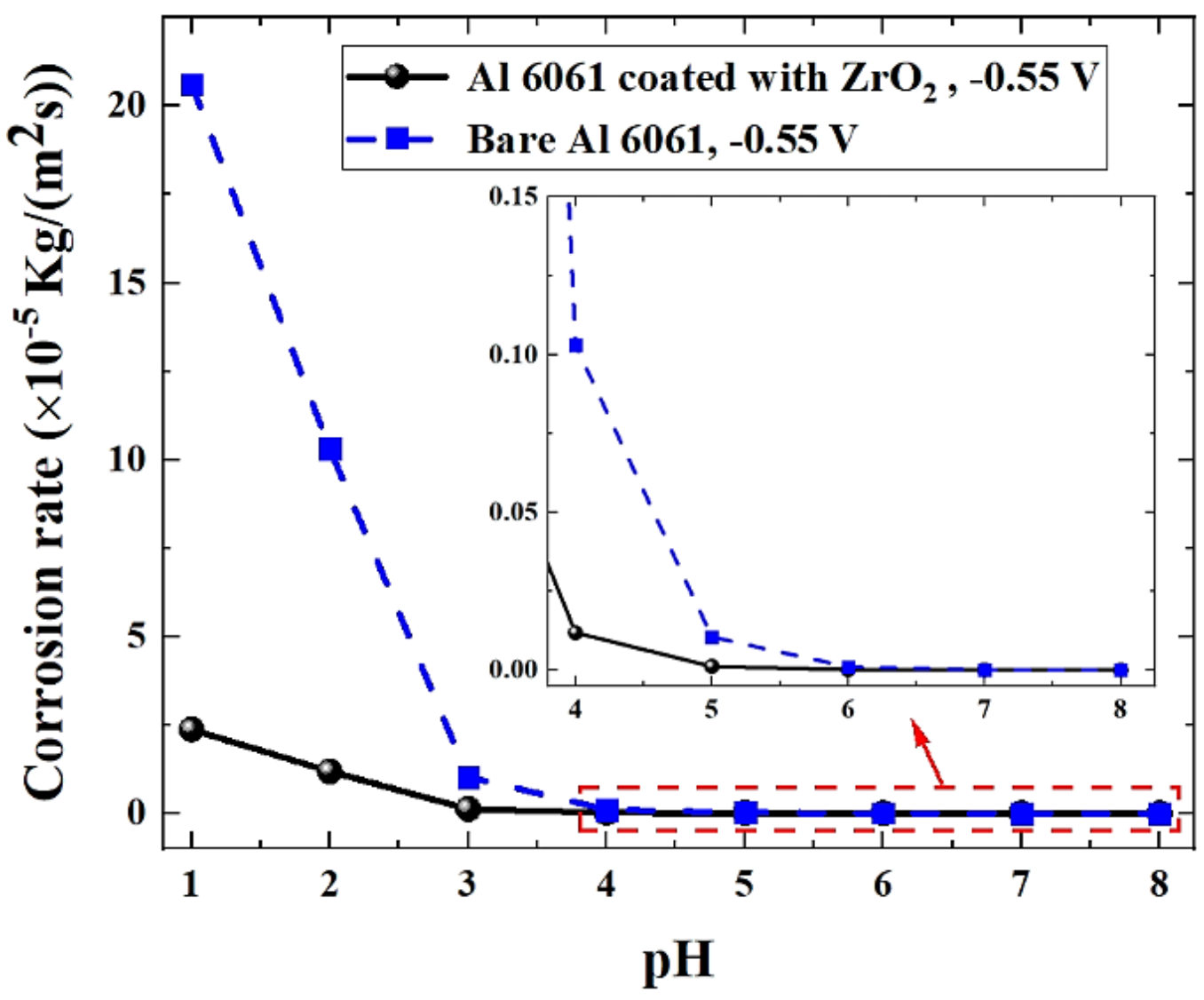

(a)

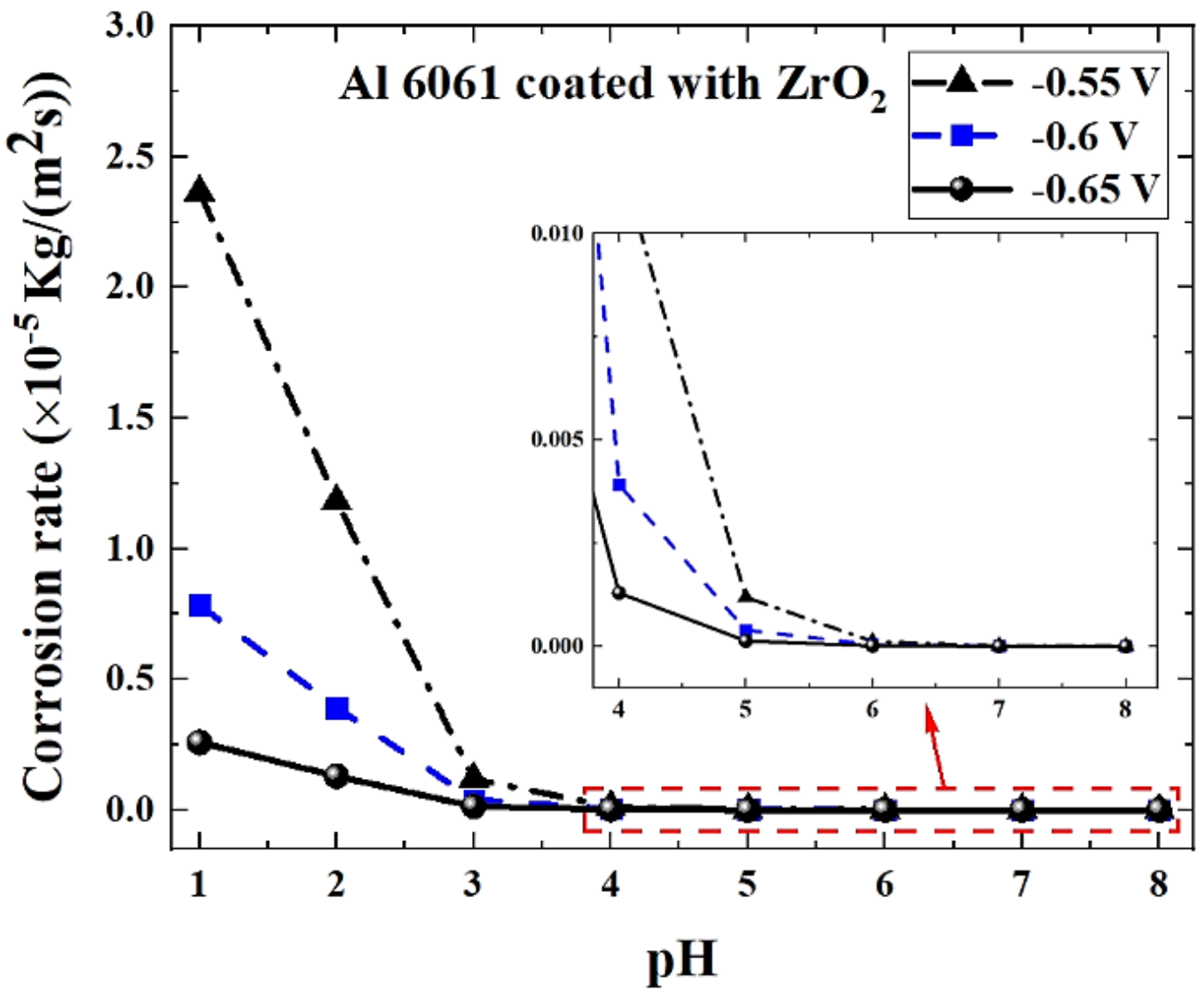

(b)

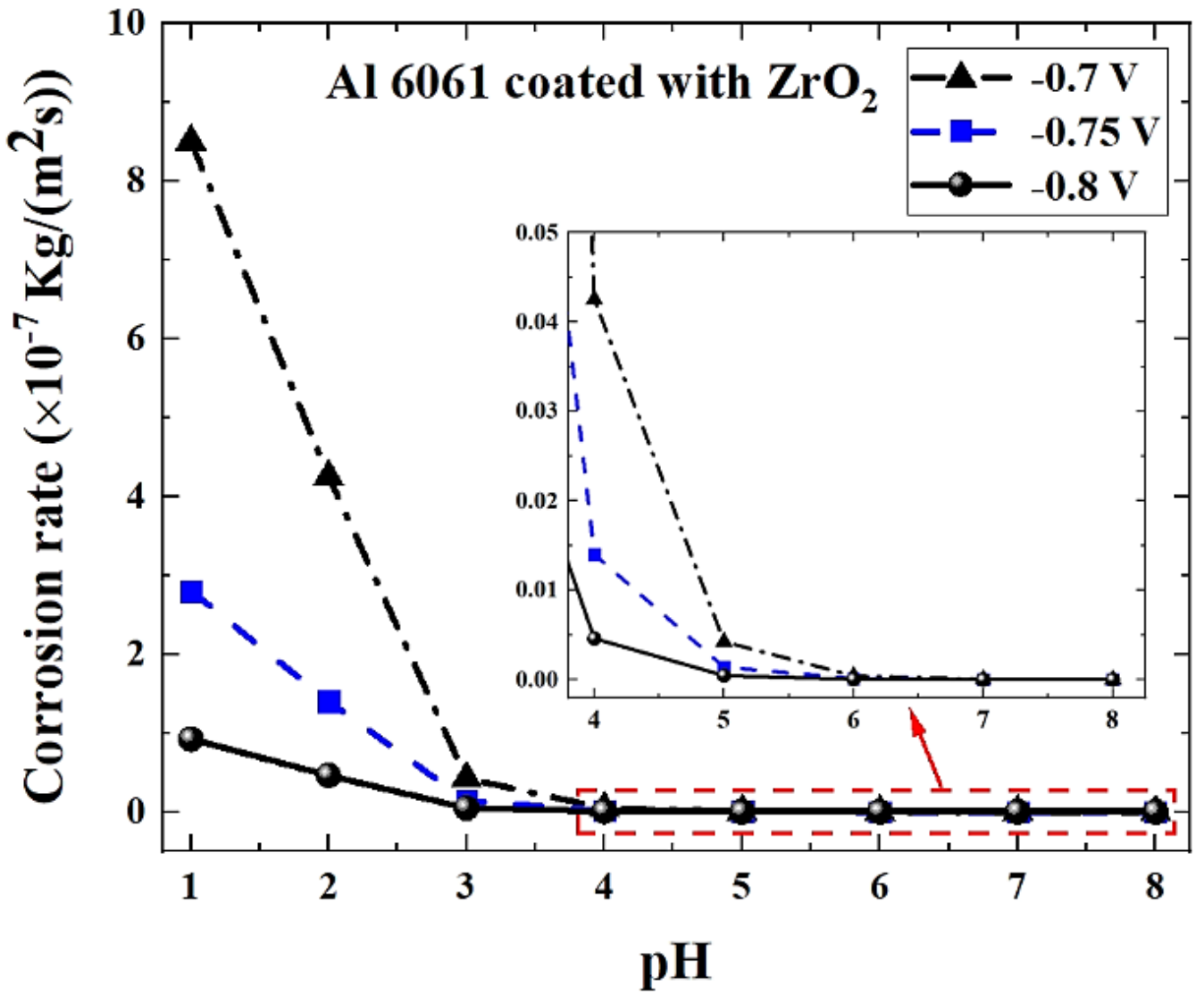

(c)

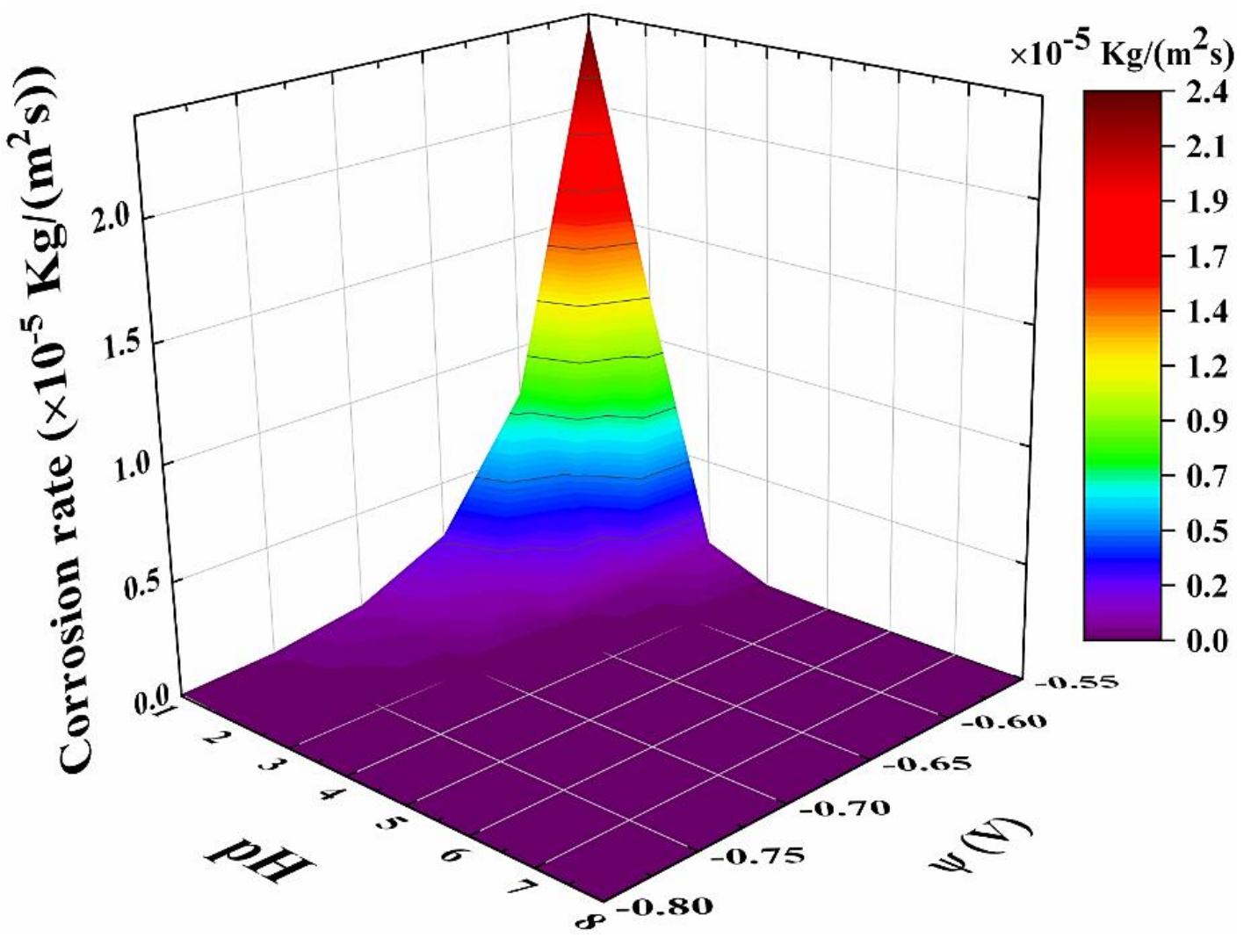

(d)

*Figure 4: The effects of pH and applied potential on the performance of localized corrosion: (a) a comparison of the corrosion rates of bare Al 6061 and Al 6061 coated with $ZrO_2$ as a function of pH at a fixed applied potential of $-0.55\ V$, (b) and (c) pH effects on the corrosion rate of Al 6061 coated with $ZrO_2$ at different applied potentials, and (d) a 3D corrosion rate-pH-applied potential diagram for Al 6061 coated with $ZrO_2$. Because of the presence of Zr near the pit surface, the corrosion rate of Al 6061 coated with $ZrO_2$ becomes nearly zero at lower pH values compared to that of bare Al 6061.*

To analyze corrosive environment within the active pit in the Al 6061 coated with $ZrO_2$, Figure 5(a-f) shows 2D contour plots of pH and electrostatic potential for different concentrations of chloride ions in the system. The initial pH for these cases was set as 7, and the pH changes with increase or decrease in chloride ion concentration as it is set for each case shown in Figure 5(a, c, e). The pH contour plots indicate that the solution aggressivity increases monotonically from the pit mouth to the pit bottom. Figure 5(a, c, e) shows that as the concentration of chloride ions increases, the electrolyte within the pit becomes more acidic (i.e., has a higher $H^+$ concentration). This event accelerates the corrosion of systems. Figure 5(b, d, f) demonstrates that the electrostatic potentials increase from the pit mouth to the pit bottom, indicating that the ohmic potential drop in the system increases. Both ohmic potential drop and solution aggressivity affect the corrosion rate of the alloy, with the former contributing negatively and the latter positively. It should be noted that the ohmic potential effect becomes negligible as the applied potential approaches the repassivation potential [4].

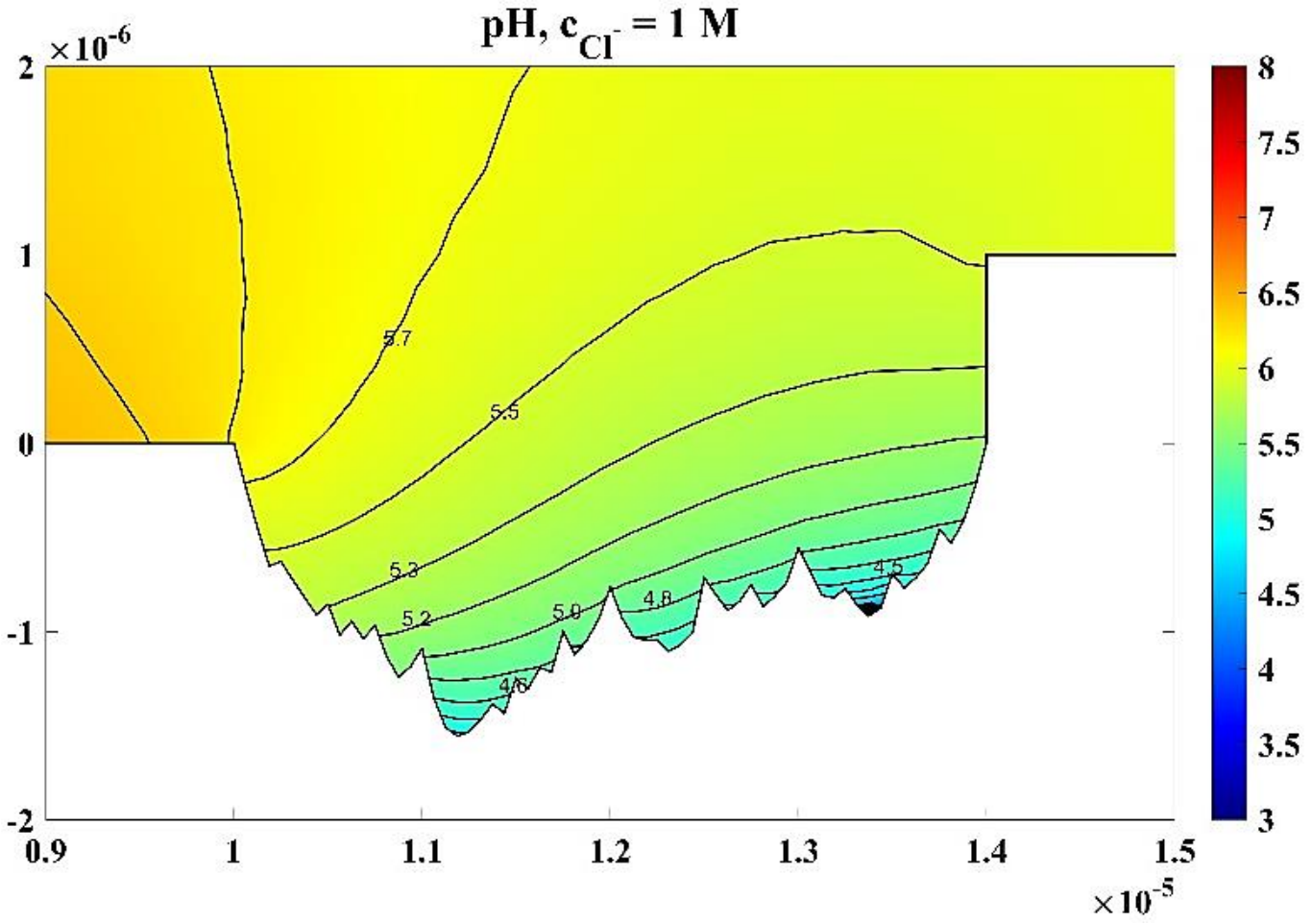

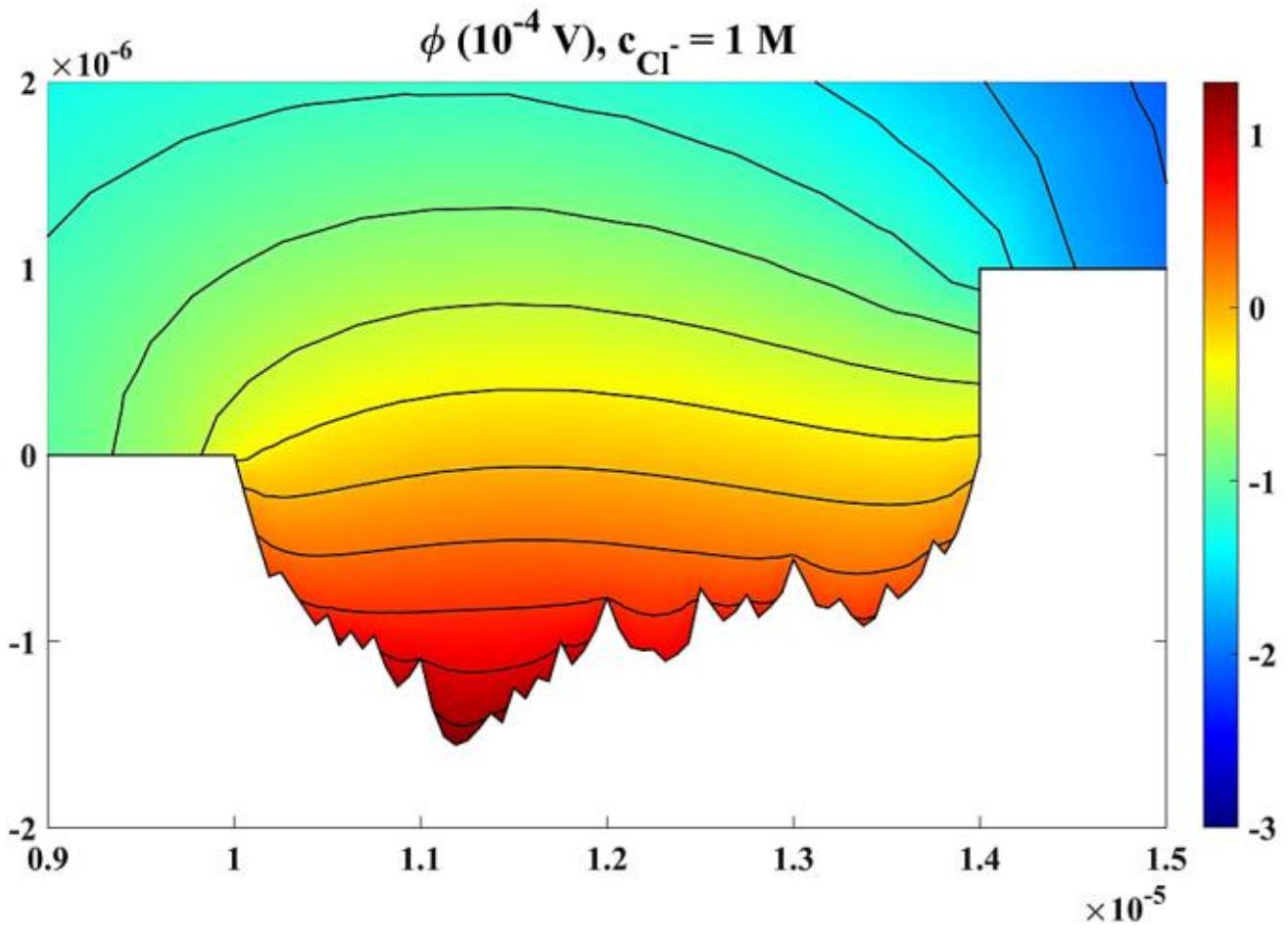

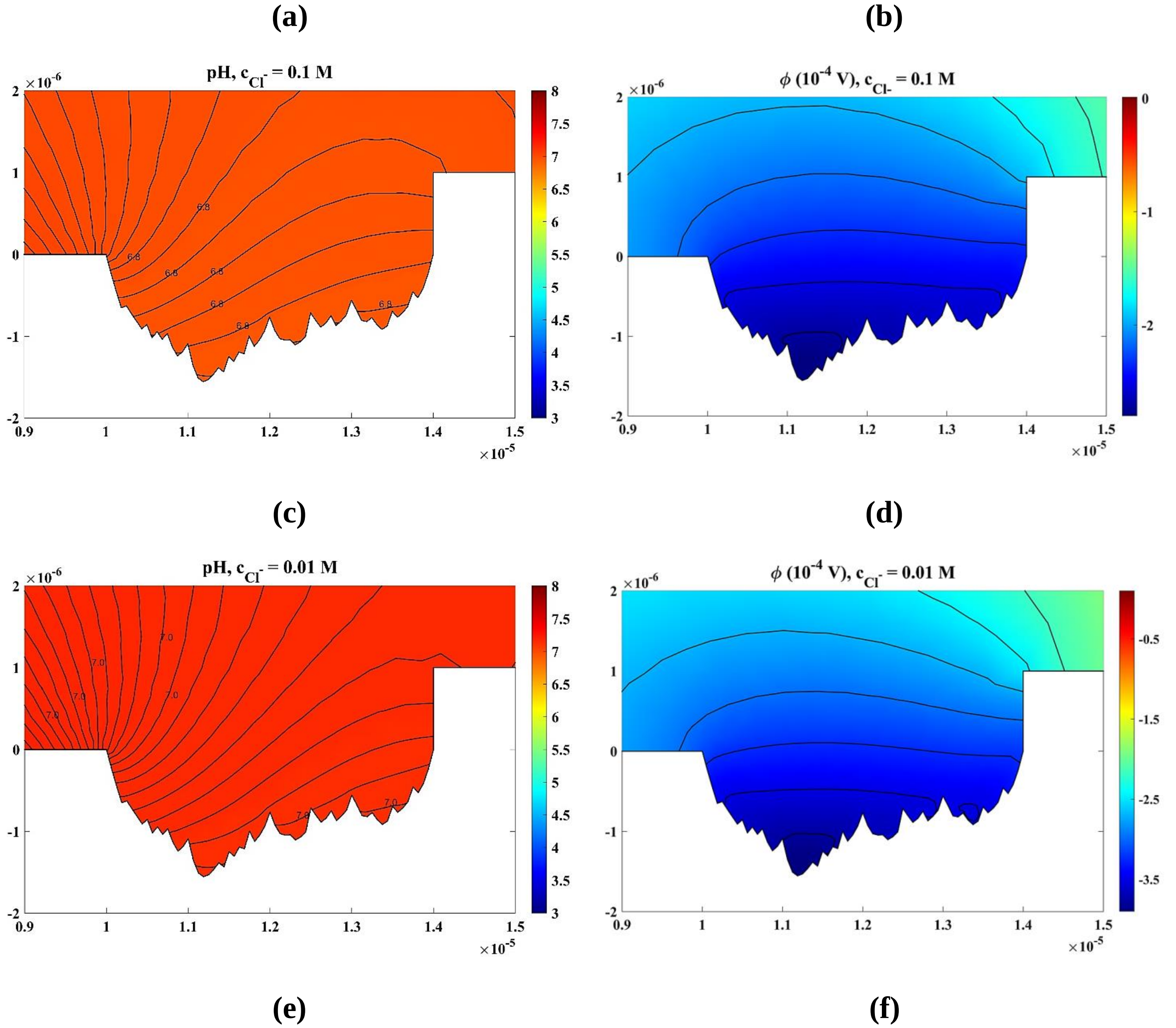

*Figure 5: Corrosive environment within an active pit of Al 6061 coated with $ZrO_2$: (a, c, e) 2D contour plots of solution pH at almost steady state, and (b, d, f) 2D contour plots of electrostatic potential. Plots are shown for three different concentrations of chloride ions: 0.01, 0.1, and 1 M. The unit of the pit dimension in r and z directions is meter. Figures indicate that ohmic and chemistry effects both contribute to the corrosion rate of a system, which must be taken into account as a coupled effect.*

### 4.3 Pit Growth Dynamics

To study the dynamics of pit growth around IMPs (θ-phase), the pit depicted in Figure 1(a) is set to propagate in a neutral sodium chloride electrolyte (0.01 M) at an applied potential of $-0.55\,V$. Figure 6 shows that the corrosion rate of both systems decreases with time throughout the corrosion process, which is consistent with experimental observations [66, 93]. The corrosion rate of Al 6061 coated with $ZrO_2$ approaches a near constant value during pit propagation much earlier than that of bare Al 6061. This is due to the presence of positively charged Al- and Zr-containing species in the pit that neutralize the charges and modify the local chemical environment close to the pit surface. The electrolyte solution becomes more aggressive over time (i.e., higher concentration of chloride ions and lower pH), and the ohmic potential drop increases in the system. Thus, the net trend of the change in corrosion rate is determined by a competition between chemistry and ohmic effects.

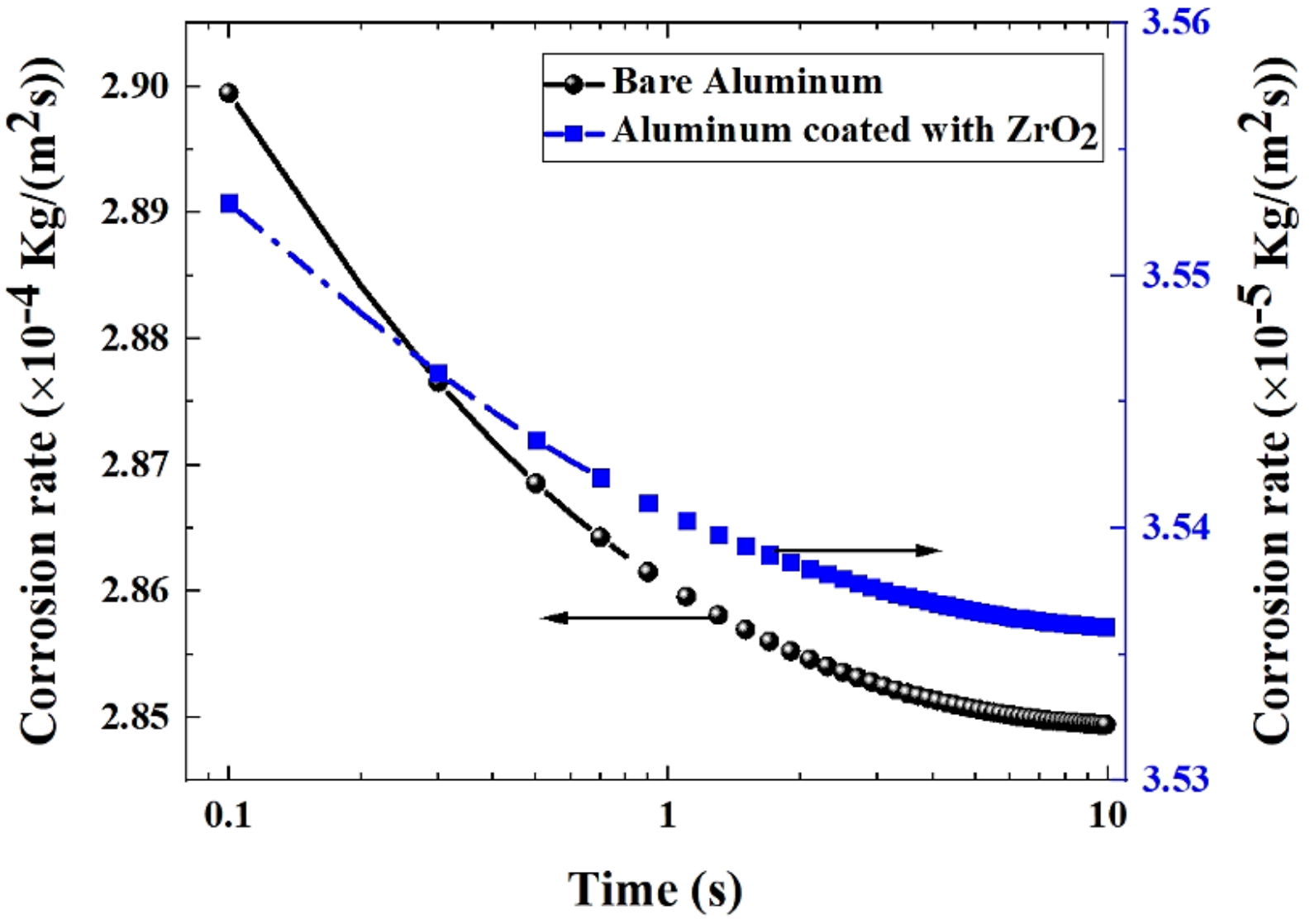

*Figure 6: Dynamics of corrosion rate during pit growth for bare Al 6061 and Al 6061 with Zr-based conversion coating.*

Figure 7(a-f) shows 2D contour plots of the growing pit on Al 6061 coated with $ZrO_2$ in the r and z – directions at different time instances. Because of the heterogeneous microstructure of aluminum alloys and moving boundary, the nature of electrochemical reactions in various locations of the pit surface was distinct and dynamically varied with time. As the pit grows in the system, more θ-phase surface is exposed. Thus, the ORR occurs on the cathodic regions (i.e., IMPs) and produces $OH^-$ ions, which neutralizes the acidic solution in the pit, resulting in a higher pH value. Consequently, the chemistry effect (i.e., solution aggressivity) on the corrosion rate of system decreases as time proceeds. The progressively exposed $Al_2O_3/ZrO_2$ protection causes the pit to expand and form an occluded geometry, which contributes to a higher potential drop within the pit (can be also understood from Figure 5(b,d,f)). As a result, while the pit grows in size, the ohmic effect on corrosion of Al 6061 coated with $ZrO_2$ becomes more pronounced. It can be concluded that the ohmic effect outweighed the chemistry effect (which was hampered by the cathodic reaction on IMPs), resulting in an overall decrease in corrosion rate during the pit propagation. Figure 7(a-f) also presents the 2D contour plots of the growing pit on Al 6061 in r and z – directions at 0.45 seconds. At 2.494 s, this instance exhibits the same pit development as the Al 6061 coated with $ZrO_2$ instance, indicating that corrosion progress is much slower in Al 6061 coated with $ZrO_2$ than in bare Al 6061.

**Aluminum 6061 Coated with $ZrO_2$**

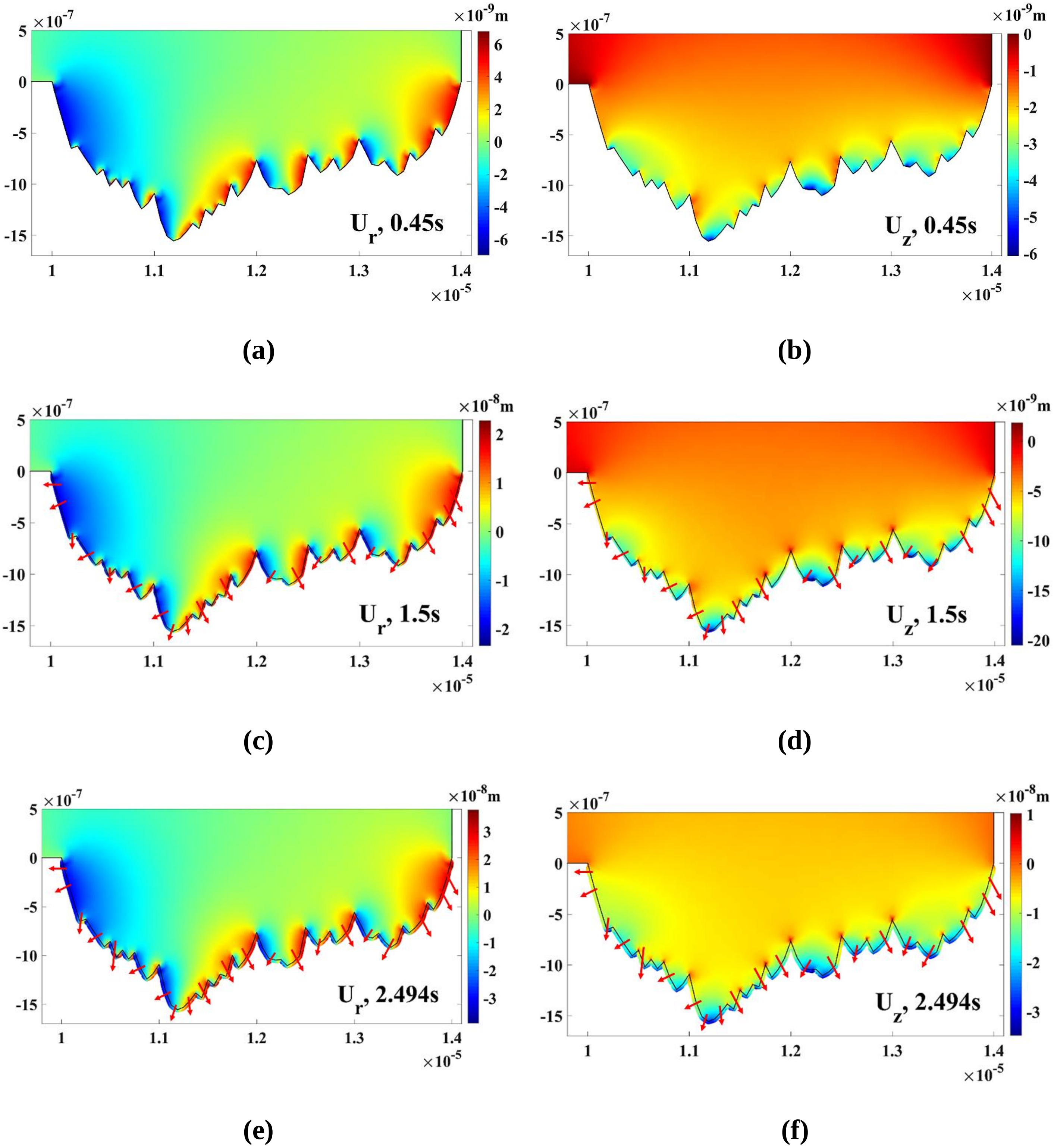

**Bare Aluminum 6061**

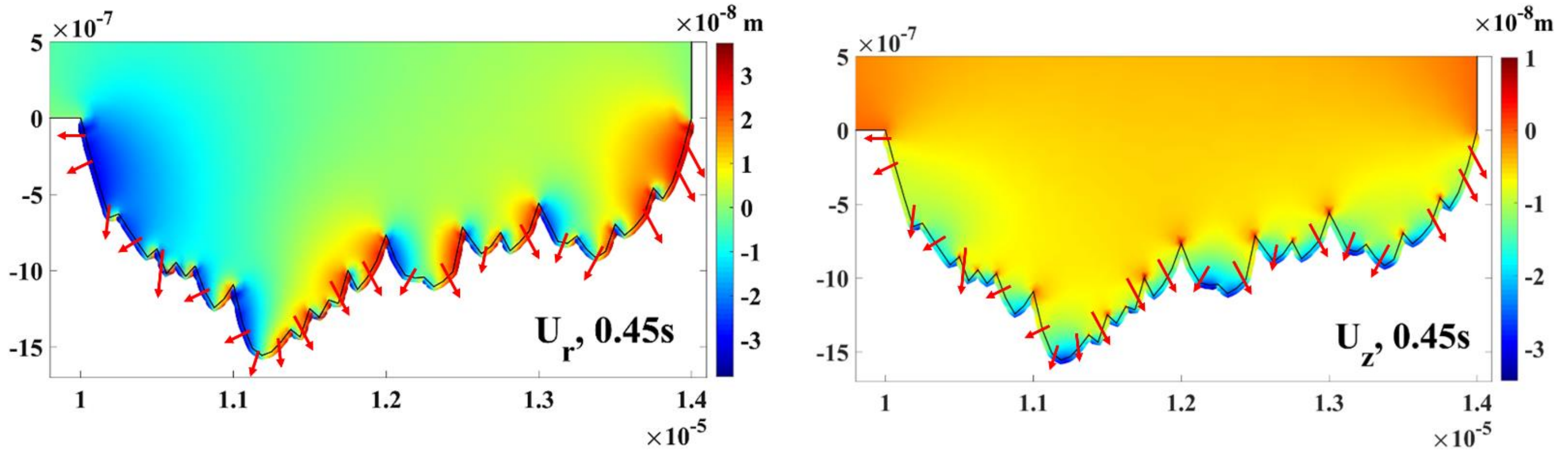

**(g)** **(h)**

*Figure 7: 2D contour plot of the growing pit in Al 6061 coated with $ZrO_2$ (a) $U_r$ at 0.45 s, (b) $U_z$ at 0.45 s, (c) $U_r$ at 1.5 s, (d) $U_z$ at 1.5 s, (e) $U_r$ at 2.494 s, and (f) $U_z$ at 2.494 s, and in bare Al 6061 (g) $U_r$ at 0.45 s and (h) $U_z$ at 0.45 s. $U_r$ and $U_z$ are the displacement in r and z directions, respectively. The unit of pit dimension in the r and z directions is meter. Comparisons show that corrosion progress in Al 6061 coated with $ZrO_2$ is much slower than in bare Al 6061. Arrows show the direction of pit growth during corrosion process.*

## 5. Conclusion

A multiscale modeling framework was developed to predict the corrosion rate and pit stability of aluminum alloys with Zr-based conversion coatings. The kinetics model, which included bulk interfacial reactions, was completed using first-principles calculations and transition-state theory. By incorporating the kinetics model into multiphysics simulations, the corrosion performance of aluminum alloys with Zr-based conversion coatings was quantified for various local corrosive environments within pits, which is very difficult to establish using *in situ* experiments. The simulation results were compared to the results of experiments reported in the literature, and they showed excellent agreements for corrosion rates of different alloys. Furthermore, good agreement was found between the DFT-TST results and experimental data for the rate constants of bulk first-order reactions occurring in the electrolyte. Later, the model was employed to quantitatively explore the effects of local corrosive environements on the corrosion performance of aluminum alloys coated with $ZrO_2$. The local corrosive fields (i.e., chemical and electrical conditions) were determined within the localized pits at different solution conditions. The following are some of the key findings from simulation results for certain electrochemical conditions as elucidated in this work: 1) a Zr-based conversion coating significantly improves the corrosion performance of aluminum alloys due to the presence of zirconium in interfacial kinetics, 2) the repassivation potential of a system (e.g., Al 6061) decreases as the concentration of chloride ions in the solution increases, leading to a less stable pit, and 3) the critical pH value for Al 6061 coated with $ZrO_2$ is lower than for bare Al 6061 because the presence of Zr changes the pit status from active to passive at lower pH.

## Data Availability

All the required data that support the findings of this study are available in the paper.

## Acknowledgments

This work was supported by the Strategic Environmental Research and Development Program (SERDP) under project WP-2742. We would like to thank the experimental team at the Center for Plasma Materials Interactions (CPMPI) of the University of Illinois at Urbana-Champaign for their sustained effort in developing these coatings using atmospheric pressure plasma and providing us with corrosion data of the components.